# High-resolution crystal structure of gelsolin domain 2 in complex with the physiological calcium ion

Michela Bollati[a,b], Emanuele Scalone[a,b], Francesco Bonì[a,b], Eloise Mastrangelo[a,b], Toni Giorgino[a,b], Mario Milani[a,b] and Matteo de Rosa[a,b]*

[a]Istituto di Biofisica, Consiglio Nazionale delle Ricerche, via Celoria 26, 20133 Milano, Italy.

[b]Dipartimento di Bioscienze, Università degli Studi di Milano, via Celoria 26, 20133 Milano, Italy.

* Correspondence to: Matteo de Rosa,

Istituto di Biofisica, Consiglio Nazionale delle Ricerche

c/o Department of BioSciences, University of Milano

Via Celoria 26, 20133 Milano, Italy

phone: +39 02 50314890 fax: +39 02 50314895

e-mail: matteo.derosa@cnr.it

# Abstract

The second domain of gelsolin (G2) hosts mutations responsible for a hereditary form of amyloidosis. The active form of gelsolin is $Ca^{2+}$-bound; it is also a dynamic protein, hence structural biologists often rely on the study of the isolated G2. However, the wild type G2 structure that have been used so far in comparative studies is bound to a crystallographic $Cd^{2+}$, in lieu of the physiological calcium. Here, we report the wild type structure of G2 in complex with $Ca^{2+}$ highlighting subtle ion-dependent differences. Previous findings on different G2 mutations are also briefly revised in light of these results.

# Keywords (6 max)

# Introduction

Gelsolin (GSN) is a multifunctional $Ca^{2+}$-dependent actin-binding protein, involved in the regulation of cytoskeleton rearrangement and cell motility by means of its severing and capping activities [1–3]. GSN is composed of six conserved copies of a prototypical domain (named G1 to G6). Gelsolin-like domains are characterized by a five-stranded β-sheet sandwiched between two α-helices [4,5]. Each GSN domain hosts at least one calcium-binding site, while only G1, G2 and G4 contains actin binding surfaces [4,5].

In $Ca^{2+}$-free conditions, gelsolin adopts a compact arrangement of the six domains [2,3,5]. Upon calcium binding, both subtle local and large global conformational changes entail the opening of at least three identifiable latches (tail latch, G1–G3 latch and G4–G6 latch) to expose the actin-binding interfaces [2,3,5,6]. Active $Ca^{2+}$-bound gelsolin is characterized by an ensemble of conformations in dynamic equilibrium till the adoption of a stable configuration upon binding to actin [4,5,7,8]. As a consequence, the crystal structure of $Ca^{2+}$-bound actin-free gelsolin has never been obtained and isolated domains have become the focus of several structural studies.

Among the six gelsolin domains, G2 has been characterized the most [9–16], because four mutations in this domain (D187N, D187Y, G167R and N184K; according to the numbering of the mature plasma protein) have been described to be responsible for a rare genetic disease named AGel amyloidosis [17–20]. AGel amyloidosis is an autosomal-dominant monogenic disease. Symptoms include corneal lattice dystrophy, cranial neuropathy, skin elasticity problems, and renal complications [21–23].

The G2-related mutations occur at different sites: D187 is part of the $Ca^{2+}$-binding site, both N and Y substitutions impair the ion binding [9,11–16]; the N184K mutant is still able to bind calcium and the geometry of the binding site matches that of wild type (WT) protein [10]; the G167R mutation was shown to promote the dimerization of GSN via a peculiar 3D domain swap mechanism [9].

Irrespective of the underlying mechanism, all the mutations so far characterized cause a drop in the thermodynamic stability of the isolated domain and increase its conformational flexibility [9–16]. As a consequence, mutated G2 becomes protease-sensitive and, in the Golgi, it is readily processed by furin, producing 8- and 5-kDa amyloidogenic peptides which self-assemble [14,24,25]. In 2018, four novel gelsolin pathogenic variants were described [26–29], which carry the mutation in other domains, either G4 or G5, and a N-terminal frame-shift.

Over the years, the superimposition of the structures of the mutated G2 domains on the WT structure has been critical in understanding the conformational changes and the destabilization caused by the mutations. The reference structure used so far hosts a crystallographic $Cd^{2+}$ in the $Ca^{2+}$ binding site instead of the physiological ion [12]. During the transition from the closed to the open conformation in WT and mutated proteins, $Ca^{2+}$ plays a crucial role. $Cd^{2+}$ and $Ca^{2+}$ have similar ionic radii (0.97 Å and 0.99 Å, respectively), however, gelsolin G2 octa-coordinates $Ca^{2+}$, while the only $Cd^{2+}$-dependent protein reported so far, carbonic anhydrase of marine diatoms, show penta-coordination [30]. In most of the organisms, these enzymes bind $Zn^{2+}$ and substitution of the ion was shown to impact the geometry of the metal binding site as well as the catalytic efficiency.

Here, we report four novel high-resolution structures of the isolated WT G2 bound to the physiological ion. Among these, the best-diffracting data is proposed as a new model for the WT protein, and differences between WT and pathological mutants are briefly reviewed in light of this result.

## Materials and methods

### Protein production

The isolated WT G2 domain (residues 151-262) was produced in *E. coli* cells and purified as previously described [9–11].

**Crystallization, structure solution and analysis**

WT G2 concentrated to 900 $\mu$M in 20 mM HEPES, pH 7.4, 100 mM NaCl and 1 mM CaCl$_2$ was used in extensive crystallization trials using an Oryx-4 robot (Douglas Instrument) and sitting-drop plates. Several different commercial screen solutions were tested, varying the protein:precipitant ratio (0.4 $\mu$l total drop volume) and the plates were stored at 20°C. Crystals of the G2 domain readily appeared in the conditions reported in Table 1, they were cryoprotected in a reservoir solution supplemented with 20% glycerol and then flash-cooled in liquid nitrogen.

X-ray diffraction data were collected at beamline I04 (Diamond Light Source, Oxfordshire) at 100°K. Data were processed using XDS [31] or DIALS [32] and scaled with AIMLESS [33]. WT G2 crystal structures were solved by molecular replacement with PHASER [34] using the G2 WT structure (PDB ID 1KCQ [12]) as a search model. Phenix refine [35] was used for the refinement of the structures while manual model building was performed with Coot [36]. All the structures reported in this manuscript were refined to commonly accepted values (see Table 1 for a complete list of data and model statistics). However, only the best-diffracting structure was deposited in the PDB with accession code 6QW3, to be proposed as the new reference model for future comparative studies. Analysis of the structures was performed with PyMOL (Schrödinger; DeLano, 2002), which was also used to prepare the figures. Global RMSD and *per residue* analysis were performed with Prosmart [37].

## Results and Discussion

### The structure of Ca$^{2+}$-bound WT G2

Owing to the difficulties associated to the study of active GSN, different constructs of the isolated G2 domain have been extensively studied and its high-resolution crystal structure solved in 2002 [12]. This relevant finding allowed the identification of the metal binding site and set the bases to dissect the pathological mechanism underlying the D187N/Y-dependent amyloidosis. Since its publication, this structural model has been used as a reference to compare the pathological variants and infer the impact of the mutations on the structure of the isolated G2. However, the crystal was obtained in the presence of high concentration of cadmium chloride and Cd$^{2+}$ was modelled in the metal binding site, instead of the physiological Ca$^{2+}$.

With the exception of G167R dimer, all the mutated G2 domains analysed so far show subtle differences compared with the WT protein. Therefore, an accurate reference model is required. In this context, extensive crystallization trials of the WT G2 in the presence of saturating Ca$^{2+}$ concentrations have been performed. Crystals readily appeared in several conditions yielding four well-diffracting (1.3-1.5 Å resolution) different crystalline forms. Table 1 shows a complete list of data collection and refinement statistics.

| Dataset | 1KCQ | 6QW3 | MP | MC | O ~~6RLK~~ |
|---|---|---|---|---|---|
| Space group | monoclinic C 2 | orthorhombic P 2 2$_1$ 2$_1$ | monoclinic primitive P 2 | monoclinic centered I 2 | orthorhombic P 2 2$_1$ 2$_1$ |
| Bound ion | Cd$^{2+}$ | Ca$^{2+}$ | Ca$^{2+}$ | Ca$^{2+}$ | Ca$^{2+}$ |

| Crystallization Conditions | 20% PEG 400, 50 mM cadmium chloride, 100 mM sodium acetate, pH 4.6 | 36% PEG 5k MME, 0.1M sodium acetate, pH 5.5 | 40% PEG 500 MME, 20% PEG 20k, 1.0M Imidazole/MES , pH 6.5 | 30% PEG 4k, 0.2M Imidazole Malate, pH 6.0 | 24% PEG 8k, 5 mM zinc acetate, 0.1M sodium cacodylate, pH 6.5 |
|---|---|---|---|---|---|
| **Data collection** | | | | | |
| Wavelength (Å) | 1.20 | 0.98 | 0.98 | 0.98 | 0.98 |
| Cell dimensions a, b, c; $\alpha$, $\beta$, $\gamma$ (Å;°) | 97, 27, 50; 90, 121, 90 | 27, 50, 71; 90, 90, 90 | 50, 27, 51; 90, 118, 90 | 50, 27, 86; 90, 92, 90 | 27, 50, 81; 90, 90, 90 |
| Unique reflections | 14,799 | 24,096 | 23,133 | 20,405 | 17,949 |
| Resolution range (Å) | 43.03-1.65 (1.74-1.65) | 70.82-1.30 (1.32-1.30) | 44.97-1.40 (1.42-1.40) | 42.82-1.45 (1.47-1.45) | 18.95-1.50 (1.54-1.50) |
| $I/\sigma(I)$ | (3.3) | 14.0 (1.8) | 16.1 (7.2) | 5.7 (0.8) | 6.5 (0.9) |
| CC 1/2 | | 1.00 (0.58) | 1.00 (0.96) | 1.00 (0.59) | 0.99 (0.29) |
| Completeness (%) | 99.2 | 99.5 (92.9) | 97.0 (93.4) | 99.9 (99.6) | 99.8 (99.9) |
| Multiplicity | 3.3 (2.7) | 10.8 (4.3) | 6.5 (5.1) | 6.5 (5.4) | 10.4 (8.7) |
| **Refinement** | | | | | |
| Resolution range (Å) | 22.25-1.65 | 41-1.30 | 44.97-1.40 | 42.25-1.45 | 18.95-1.50 |
| Rwork/Rfree* | 0.177/0.233 | 0.154/0.174 | 0.137/0.178 | 0.177/0.219 | 0.204/0.241 |

| RMSD Bonds/angles (Å/°) | 0.007/1.900 | 0.016/1.469 | 0.016/1.324 | 0.006/0.856 | 0.015/1.902 |
|---|---|---|---|---|---|
| Ramachandran outliers (%) | 0 | 0 | 0 | 0 | 0.8 |
| B factors (Å²)§ | 22 | 21 | 19 | 28 | 23 |

**Table 1: Data collection and refinement statistics of WT G2s bound to Ca²⁺.** * $R_{work} = \Sigma_{hkl} ||Fo| - |Fc|| / \Sigma_{hkl} |Fo|$ for all data, except 5-10%, which were used for $R_{free}$ calculation. § Average temperature factors over the whole structure.

One of these crystals (O) grew in 5 mM zinc acetate; however, no anomalous signal was detected, confirming the presence of $Ca^{2+}$ in the metal binding site. All four models should be considered equally representative of WT $Ca^{2+}$-bound G2. Comparison of these models facilitates the identification of structural differences caused by the crystalline packing or by some intrinsic flexibility of specific stretches of the protein. Overall the four structures superimpose particularly well (Figure 1A and Table 2) and minor deviations of the backbone are observed only in the loop connecting β5 to α2 (residues 236-240). Owing to the quality of the data, one of the two orthorhombic crystals was deposited in the PDB with accession code 6QW3 and represents the reference structure chosen for the comparison of the WT G2 $Ca^{2+}$-bound form with the $Cd^{2+}$-bound form (Figure 1B). 6QW3 was subjected to the RMSD *per* residue analysis versus all the others WT G2-$Ca^{2+}$ complex structures; the values were averaged to calculate an unbiased value (i.e. a value independent of crystal packing) that was subtracted from the RMSD of 6QW3 *Vs.* 1KCQ (Figure 1C).

This analysis should specifically mark differences due to the nature of the bound ion. Indeed, only two regions of G2 show significant differences in RMSD values: i) the hinge loop (hosting the furin aberrant cleavage site, residues 168 and 169) and ii) the C-terminal tail of the domain (residues 255-260).

Differences in the hinge loop might be due to mis-modelling of a noisy area of the structure (Figure 3). Contrarily, the electron density of the C-terminus is of sufficient quality to take into consideration a proper conformational change (Figure 2). In addition to the opportunistic ion, three additional $Cd^{2+}$ were modelled in the 1KCQ asymmetric unit. One of these $Cd^{2+}$ is relevant for our analysis as it is coordinated by the C-terminal residue E258, less than 6 Å afar from the physiological metal site (Figure 2). It is difficult to determine the impact of this $Cd^{2+}$ on the structure of the G2, because it sits near a crystallographic two fold axis (0.4 Å) and it has been refined to 0.5 occupancy. However, we cannot exclude that this second $Cd^{2+}$ contributes to the subtle differences observed in the C-terminal region.

We can conclude that, to allocate the $Cd^{2+}$ ion in the metal binding site, the C-terminal tail undergoes subtle but significant rearrangements. These adjustment of the backbone are related to an increased flexibility of the hinge loop.

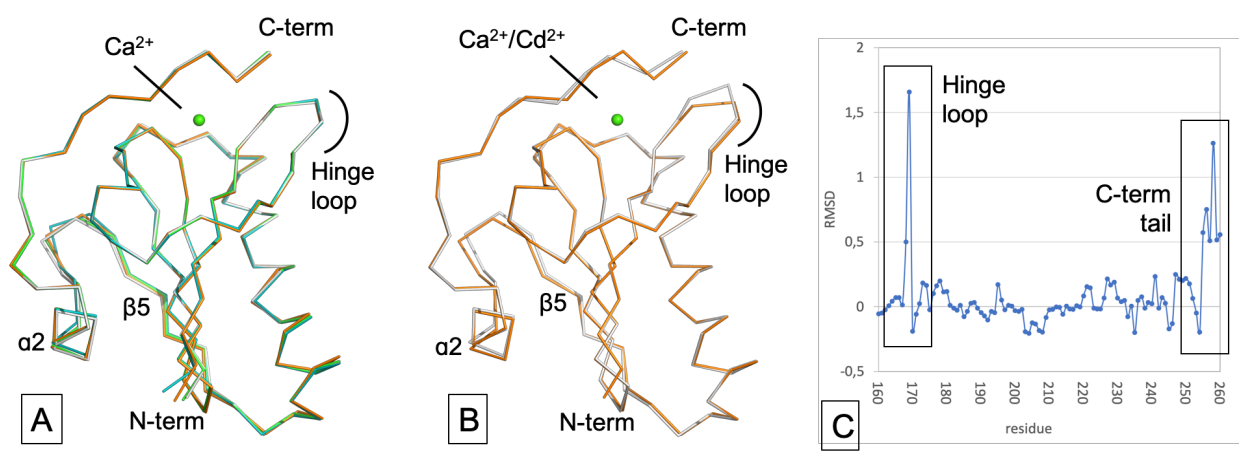

*Figure 1: Ion-dependent differences in the WT G2. A)* Cα-traces of the 4 aligned structures of G2 WT bound to Ca$^{2+}$ (6QW3 in orange, MP in cyan, MC in green and O in white). *B)* Cα-traces of Ca$^{2+}$-bound 6QW3 (in orange) superimposed on the Cd$^{2+}$-bound 1KCQ (in grey) *C)* "Net" RMSD values of 6QW3 Vs. 1KCQ (Å) (i.e. RMSD subtracted from average RMSD Vs. other Ca$^{2+}$-bound structures).

|  | **6QW3** | MP | MC | O |
|---|---|---|---|---|
| **1KCQ** | 0.537 | 0.551 | 0.493 | 0.611 |
| **6QW3** |  | 0.461 | 0.260 | 0.423 |
| MP |  |  | 0.295 | 0.347 |
| MC |  |  |  | 0.287 |

*Table 2: Ion-dependent differences in the WT G2.* Multiple alignment scores were obtained from ProSMART (RMSD values, Å); residues 159-260 (102 Cα atoms) were used in this analysis.

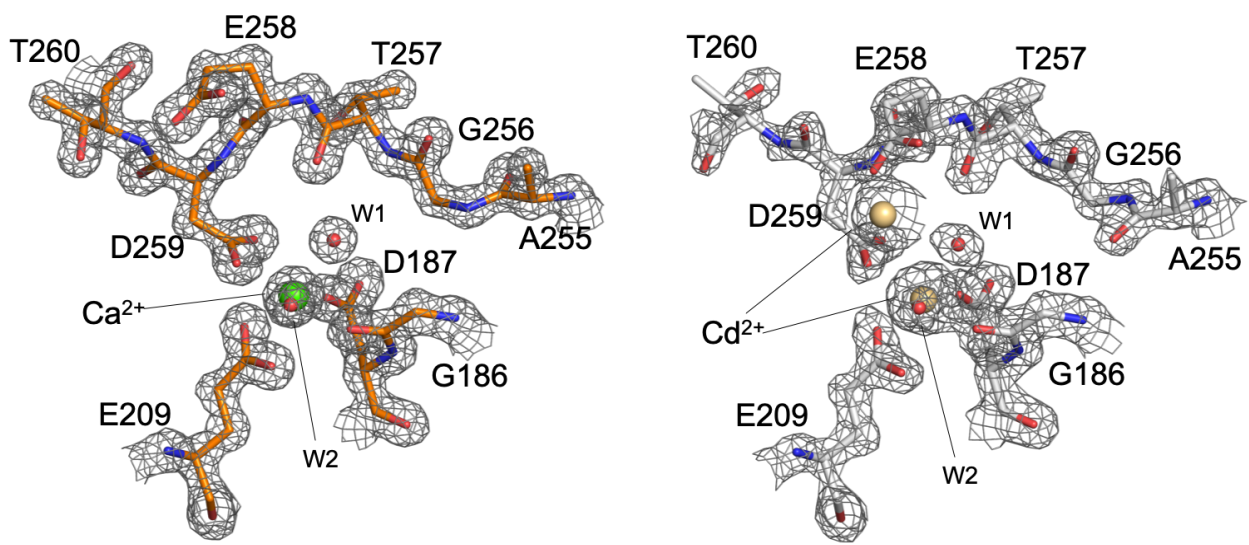

**Figure 2: Details of the metal binding site in the WT G2 structure.** *Left: sticks model of WT G2 (orange) bound to $Ca^{2+}$ (green sphere). Right: 1KCQ (grey), bound to $Cd^{2+}$ (yellow). Electron density is contoured at 1.5 σ.*

## Structural differences between $Ca^{2+}$-bound WT and mutated G2

Different arrangements of the C-terminal tail are observed in most of the G2 domain pathological variants. Two crystal structures of N184K variant were obtained and in one of the resulting models the terminal tail was fully displaced, leaving the calcium binding pocket exposed to solvent [10]. Although this conformation was partly due to the stabilization induced by the crystal packing, this behavior might result from the increased flexibility of the C-terminus. In the G167R variant, which has been crystallized only in the domain-swapped conformation, the terminal tail is displaced, with the metal binding site accessible to the solvent and a citrate molecule completing the coordination of the ion [9]. The structure of the D187N variant, although obtained as a complex with a stabilizing nanobody, showed a very flexible C-terminus that could only be modelled to a limited extent [11].

In conclusion, in all the variants structurally characterized so far, major differences have been observed between WT and mutants' C-terminal dynamics. The subtle differences observed between the $Ca^{2+}$- and $Cd^{2+}$-bound WT are not functionally significant. Moreover, the study of the mechanism of action of the chaperoning nanobody revealed that the destabilization of the C-terminal tail is not a major determinant of mutated gelsolin susceptibility to aberrant proteolysis [11]. In fact, the nanobody is able to restore thermodynamic stability of mutated G2 without recovering the native conformation of the C-terminus.

The impact of the pathological mutations may have been so far underestimated because the same region in the WT protein has been considered intrinsically flexible. $Cd^{2+}$ binding instead of the physiological ion leads to a destabilization of the hinge loop, in particular of residues 168 and 169, which becomes poorly resolved (Figure 3). Low quality of the electron density was observed in other destabilized conformation of the protein, such as the N184K variant which, despite the atomic resolution shows poorly resolved residues 168 and 169 (Figure 3).

Contrary, in the stable or stabilized (by nanobody) variants, the residues belonging to the loop could be unambiguously traced, as it is the case of D187N and G167R. This observation unravels some controversies between previously reported experimental and computational data. MD simulations have been extensively used to characterise the dynamic behaviour of gelsolin mutants and often highlighted an increased flexibility of the hinge loop in the destabilized mutants. D187N showed significant differences with respect to the WT when the simulation-derived B-factors were compared [11,12]. Similar results were reported for MD simulations on the monomeric form of the G167R variant [9].

In conclusion, our results on G2 WT in complex with $Ca^{2+}$ demonstrate that the increased flexibility of the hinge loop is a major determinant of mutated G2 instability. The aberrant cleavage of the hinge loop by furin is the first step of the pathological pathway which leads to mutated gelsolin deposition. In previous studies, pathogenic mutations have been described as responsible for the *exposure* of the furin recognition site. However, in the open conformation of gelsolin, the hinge loop is likely solvent accessible. We already showed that the C-terminus of G2 provides only a minor steric protection, i.e. its destabilization is required but not sufficient [10,11]. A rearrangement of the hinge loop is therefore likely needed to fit into the active site of the protease that has been shown to recognize rather stretched peptides [38,39].

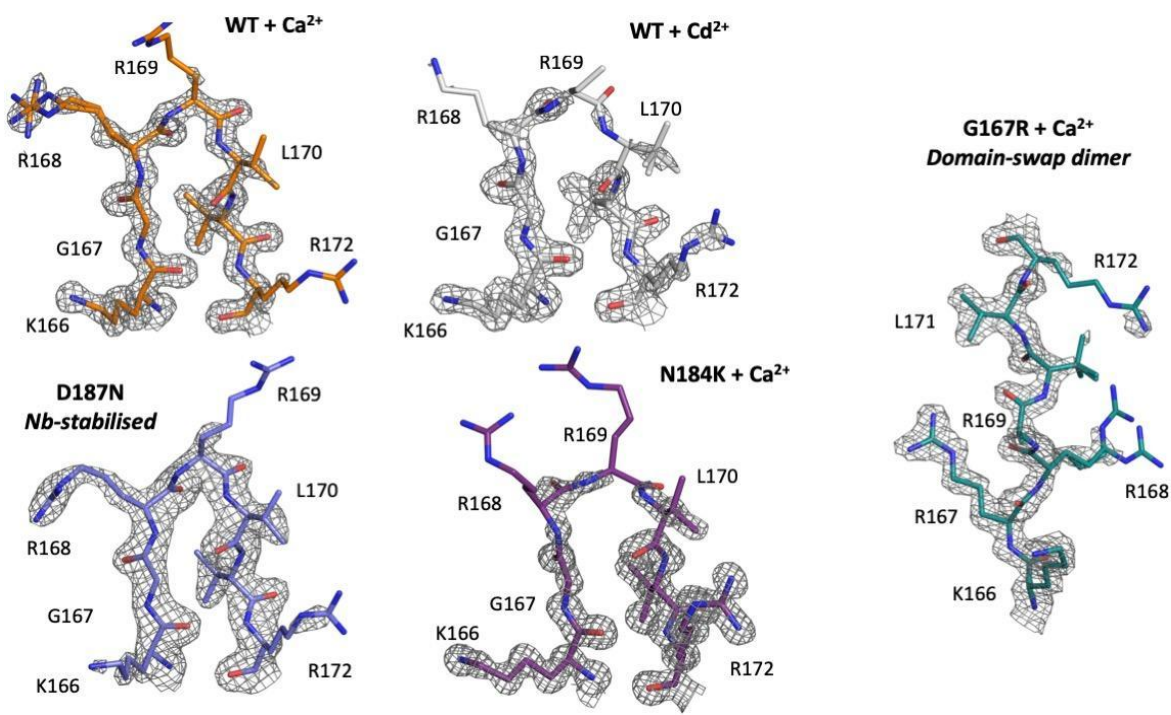

***Figure 3: Impact of ions and mutations on the conformation and flexibility of the hinge loop.*** *Close-view of the hinge loop (residues 166 to 172) in the available G2 structures: WT in complex with $Ca^{2+}$ (WT+$Ca^{2+}$, PDB ID 6QW3, this work) or with $Cd^{2+}$ (WT+$Cd^{2+}$, PDB ID 1KCQ [12]), D187N in complex with the chaperone nanobody (D187N Nb-stabilized, PDB ID 6H1F [11]), N184K in complex with $Ca^{2+}$ (N184K+$Ca^{2+}$, PDB ID 5FAF [10]) and G167R in the domain-swapped dimeric form bound to $Ca^{2+}$*

(G167R+Ca$^{2+}$, PDB ID 5O2Z [9]). residues are represented as sticks and electron density displayed at 1.5 σ.

## Acknowledgments

This research was supported by the Amyloidosis Foundations (Michigan, United States) thanks to a Research Grant awarded to MdR. The diffraction experiments were performed on I04 at Diamond Light Source (DLS, United Kingdom). We are thankful for the provided beamtime and assistance during the measurements.